\documentclass{elsart}

\usepackage{indentfirst}
\usepackage{graphicx}
\usepackage{psfrag}

\usepackage{subfigure}

\usepackage[english]{babel}
\usepackage[latin1]{inputenc}

\usepackage[sort,sectionbib]{natbib}

\usepackage{xspace} 
\usepackage{amsmath,amssymb}

\usepackage{color}
\definecolor{oneblue}{rgb}{0,0.0,0.75}
\usepackage[colorlinks,
            urlcolor=oneblue,
            linkcolor=oneblue,
            citecolor=oneblue,
            bookmarksopen=false,
            pagebackref]{hyperref}

\newcommand{\od}[2]{\frac{d#1}{d#2}}
\newcommand{\pd}[2]{\frac{\partial#1}{\partial#2}}
\newcommand{\abs}[1]{\left|#1\right|}

\newcommand{\m}{\mu^2}

\newcommand{\sech}{{\rm sech}\,}

\renewcommand{\Im}{\mathop{\mathrm{Im}}}

\def\g{\vec{g}}
\def\v{\vec{v}}
\def\u{\vec{u}}

\def\0{\vec{0}}
\def\n{\vec{n}}
\def\x{\vec{x}}

\def\vpsi{\vec{\psi}}

\def\T{\mathcal{T}}
\def\A{\mathcal{A}}
\def\E{\mathcal{E}}
\def\R{\mathbb{R}}

\def\O{\mathcal{O}}

\def\M{\mathbf{M}}
\def\grad{\nabla}
\def\div{\nabla\cdot}

\def\eps{\varepsilon}
\def\ssigma{\underline{\underline{\vec{\sigma}}}}
\def\ttau{\underline{\underline{\tau}}}

\def\I{\mathbb{I}}

\journal{Eur. J. Mech. B/Fluids}

\begin{document}

\begin{frontmatter}

\title{Visco-potential free-surface flows and long wave modelling}
\author[Dutykh]{Denys Dutykh\corauthref{cor}}
\address[Dutykh]{CMLA\thanksref{leave}, ENS Cachan, CNRS, PRES UniverSud, 61 Av. President Wilson, F-94230 Cachan, France}
\thanks[leave]{Now at LAMA, University of Savoie,
CNRS, Campus Scientifique, 73376 Le Bourget-du-Lac Cedex, France}

\corauth[cor]{Corresponding author.}
\ead{Denys.Dutykh@univ-savoie.fr}

\begin{abstract}
In a previous study \cite{DutykhDias2007} we presented a novel visco-potential free surface flows formulation. The governing equations contain local and nonlocal dissipative terms. From physical point of view, local dissipation terms come from molecular viscosity but in practical computations, rather eddy viscosity should be used. On the other hand, nonlocal dissipative term represents a correction due to the presence of a bottom boundary layer. Using the standard procedure of Boussinesq equations derivation, we come to nonlocal long wave equations. In this article we analyse dispersion relation properties of proposed models. The effect of nonlocal term on solitary and linear progressive waves attenuation is investigated. Finally, we present some computations with viscous Boussinesq equations solved by a Fourier type spectral method.
\end{abstract}


\end{frontmatter}


\section{Introduction}\label{sec:intro}

Even though the irrotational theory of free-surface flows can predict successfully many observed wave phenomena, viscous effects cannot be neglected under certain circumstances. Indeed the question of dissipation in potential flows of fluid with a free surface is an important one. As stated by \cite{LH1992}, it would be convenient to have equations and boundary conditions of comparable simplicity as for undamped free-surface flows. The peculiarity here lies in the fact that the viscous term in the Navier--Stokes (NS) equations is identically equal to zero for a velocity deriving from a potential. There is also a problem with boundary conditions. It is well known that the usual non-slip condition on the solid boundaries does not allow to simulate free surface flows in a finite container. Hence, some further modifications are required to permit the free surface particles to slide along the solid boundary. These difficulties were overcome in our recent work \cite{DutykhDias2007} and \cite{Dias2007}.

The effects of viscosity on gravity waves have been addressed since the end of the nineteenth century in the context of the linearized Navier--Stokes (NS) equations. It is well-known that Lamb \cite{Lamb1932} studied this question in the case of oscillatory waves on deep water. What is less known is that Boussinesq studied this effect as well \cite{Boussinesq1895}.
In this particular case they both showed that
\begin{equation}\label{eq:classical0}
  \od{\alpha}{t} = -2\nu k^2 \alpha,
\end{equation}
where $\alpha$ denotes the wave amplitude, $\nu$ the kinematic viscosity of the fluid and $k = 2\pi/\ell$ the wavenumber of the decaying wave. Here $\ell$ stands for the wavelength. This equation leads to the classical law for viscous decay, namely
\begin{equation}\label{eq:expdecay}
	\alpha(t)=\alpha_0 e^{-2\nu k^2 t}.
\end{equation}
Let us consider a simple numerical application with $g = 9.8$ $m/s^2$, $\ell = 3$ $m$ and molecular viscosity $\nu = 10^{-6}$ $m^2/s$. According to the formula (\ref{eq:expdecay}), this wave will take $t_0 \approx 8\times 10^4$ $s$ or about one day before losing one half of its amplitude. This wave will attain velocity
\begin{equation*}
  c_p = \sqrt{\frac{g}{k}} \approx 2.16\; m/s
\end{equation*}
and travel the distance equal to $L = c_p\cdot t_0 \approx 170$ $km$. This estimation is exaggerated since the classical result of Boussinesq and Lamb does not take into account energy dissipation in the bottom boundary layer. We will discuss the question of linear progressive waves attenuation in Section \ref{sec:damping}. Another point is that the molecular viscosity $\nu$ should be replaced by eddy viscosity $\nu_t$ which is more appropriate in most practical situations (see Remark \ref{rem:eddy} for more details).

The importance of viscous effects for water waves has been observed in various experimental studies. For example, in \cite{Zabusky1971} one can read
\begin{quote}
 \dots However, the amplitude disagrees somewhat, and we suppose that this might be due to the viscous dissipation\dots
\end{quote}
\cite{Wu1981} also mentions this drawback of the classical water wave theory:
\begin{quote}
  \dots the peak amplitudes observed in the experiments are slightly smaller than those predicted by the theory. This discrepancy	can be ascribed to the neglect of the viscous effects in the theory\dots
\end{quote}
Another example is the conclusion of \cite{Bona1981}:
\begin{quote}
  \dots it was found that the inclusion of a dissipative term was much more important than the inclusion of the nonlinear term, although the inclusion of the nonlinear term was undoubtedly beneficial in describing the observations\dots
\end{quote}

Another source of dissipation is due to bottom friction. An accurate computation of the bottom shear stress $\vec{\tau}$ is crucial for calculating sediment transport fluxes. Consequently, the predicted morphological changes will greatly depend on the chosen shear stress model. Traditionally, this quantity is modelled by a Chézy-type law
\begin{equation*}
  \left.\vec{\tau}\right|_{z=-h} = C_f\rho|\u_h|\u_h,
\end{equation*}
where $\u_h = \u(x,y,-h,t)$ is the fluid velocity at the bottom, $C_f$ is the friction coefficient and $\rho$ is the fluid density. Two other often used laws can be found in \cite[Section 4.3]{Dutykh2007}, for example. One problem with this model is that $\vec{\tau}$ and $\u_h$ are in phase. It is well known \cite{Liu2006} that in the case of a laminar boundary layer, the bottom stress $\left.\vec{\tau}\right|_{z=-h}$ is $\frac{\pi}{4}$ out of phase with respect to the bottom velocity.

Water wave energy can be dissipated by different physical mechanisms. The research community agrees at least on one point: the molecular viscosity is unimportant. Now let us discuss more debatable statements. For example if we take a tsunami wave and estimate its Reynolds number, we find $Re\approx 10^6$. So, the flow is clearly turbulent and in practice it can be modelled by various eddy viscosity models. On the other hand, in laboratory experiments the  Reynolds number is much more moderate and sometimes we can neglect this effect. When nonbreaking waves feel the bottom, the most efficient mechanism of energy dissipation is the bottom boundary layer. This is the focus of our paper. We briefly discuss the free surface boundary layer and explain why we do not take it into account in this study. Finally, the most important (and the most challenging) mechanism of energy dissipation is wave breaking. This process is extremely difficult from the mathematical but also the physical and numerical points of view since we have to deal with multivalued functions, topological changes in the flow and complex turbulent mixing processes. Nowadays the practitioners can only be happy to model roughly this process by adding ad-hoc dissipative terms when the wave becomes steep enough.

In this work we keep the features of undamped free-surface flows while adding dissipative effects. The classical theory of viscous potential flows is based on pressure and boundary conditions corrections \cite{Joseph2004} due to the presence of viscous stresses. We present here another approach.

Currently, potential flows with ad-hoc dissipative terms are used for example in direct numerical simulations of weak turbulence of gravity waves \cite{Dyachenko2003,Dyachenko2004,Zakharov2005}. There have also been several attempts to introduce dissipative effects into long wave modelling \cite{Mei1994,Dutykh2007,CG,Khabakhpashev1997}. We would like to underline that the last paper \cite{Khabakhpashev1997} contains also a nonlocal dissipative term in time.

The present work is a direct continuation of the recent studies \cite{Dias2007, DutykhDias2007}. In \cite{Dias2007} the authors considered periodic waves in infinite depth and derivation was done in two-dimensional (2D) case, while in \cite{DutykhDias2007} we removed these two hypotheses and all the computations are done in 3D. This point is important since the vorticity structure is more complicated in 3D. In other words we considered a general wavetrain on the free surface of a fluid layer of finite depth. As a result we obtained a formulation which contains a nonlocal term in the bottom kinematic condition. Later we discovered that the nonlocal term in exactly the same form was derived in \cite{Liu2004}. The inclusion of this term is natural since it represents the correction to potential flow due to the presence of a boundary layer. Moreover, this term is predominant since its magnitude scales with $\O(\sqrt{\nu})$, while other terms in the free-surface boundary conditions are of order $\O(\nu)$. The importance of this effect was pointed out in the classical literature on the subject \cite{Lighthill1978}:
\begin{quote}
  \dots Bottom friction is the most important wherever the water depth is substantially less than a wavelength so that the waves induce significant horizontal motions near the bottom; the associated energy dissipation takes place in a boundary layer between them and the solid bottom\dots
\end{quote}
This quotation means that this type of phenomenon is particularly important for shallow water waves like tsunamis, for example \cite{Dias2006, Dutykh2007a}. Here we present several numerical computations based on the newly derived governing equations and analyse dispersion relation properties.

We would like to mention here a paper of N. Sugimoto \cite{Sugimoto1991}. The author considered initial-value problems for the Burgers equation with the inclusion of a hereditary integral known as the fractional derivative of order $\frac12$. The form of this term was justified in previous works \cite{Sugimoto1989, Sugimoto1990}. Note, that from fractional calculus point of view our nonlocal term (\ref{fond_dissip}) is also a half-order integral.

Other researchers have obtained nonlocal corrections but they differ from ours \cite{KM}. This discrepancy can be explained by a different scaling chosen by Kakutani \& Matsuuchi in the boundary layer. Consequently, their governing equations contain a nonlocal term in space. The performance of the present nonlocal term (\ref{fond_dissip}) was studied in \cite{Liu2006}. The authors carried out in a wave tank a set of experiments, analyzing the damping and shoaling of solitary waves. It is shown that the viscous damping due to the bottom boundary layer is well represented by the nonlocal term. Their numerical results fit well with the experiments. The model not only properly predicts the wave height at a given point but also provides a good representation of the changes on the shape and celerity of the soliton. We can conclude that the experimental study by P. Liu et al. \cite{Liu2006} validates this theory.

The present article is organized as follows. In Section \ref{sec:anatomy} we estimate the rate of viscous dissipation in different regions of the fluid domain. Then, we present basic ideas of derivation and come up with visco-potential free-surface flows formulation. At the end of Section \ref{sec:derivation} we give corresponding long wave models: nonlocal Boussinesq and KdV equations. Section \ref{sec:disprel} is completely devoted to the analysis of linear dispersion relation of complete and long wave models introduced in previous section. Last two sections deal with linear progressive and solitary waves attenuation respectively. Finally, the paper is ended by some conclusions and perspectives.


\section{Anatomy of dissipation}\label{sec:anatomy}

\begin{figure}[htbp]
\centering
\psfrag{O}{$O$}
\psfrag{x}{$\x$}
\psfrag{z}{$z$}
\psfrag{S}{$S_f$}
\psfrag{T}{$S_b$}
\psfrag{A}{$R_f$}
\psfrag{B}{$R_b$}
\psfrag{R}{$R_i$}
\includegraphics[width=10cm]{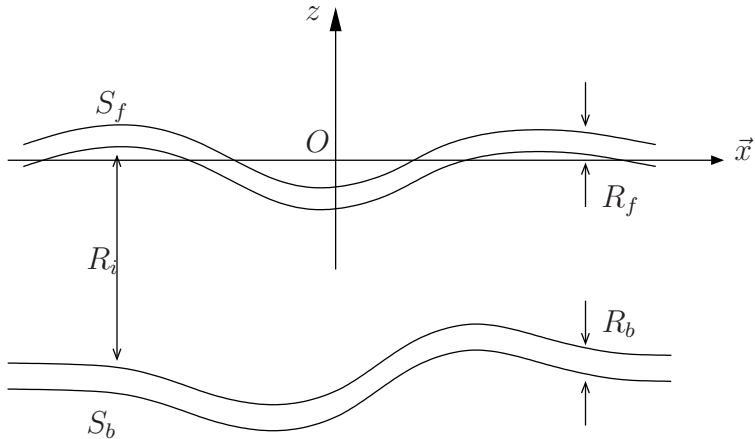}
\caption[Conventional partition of the flow region]{Conventional partition of the flow region into interior region and free surface, bottom boundary layers.}
\label{fig:anatomy}
\end{figure}

In this section we briefly discuss the contribution of different flow regions into water wave energy dissipation. We conventionally \cite{Mei1994} divide the flow into three regions illustrated on \figurename~\ref{fig:anatomy}. On this figure $S_f$ and $S_b$ stand for free surface and bottom respectively. Then, $R_i$, $R_f$ and $R_b$ denote the interior region, free surface and bottom boundary layers.

In order to make some estimates we introduce the notation which will be used in this section: $\mu$ is the dynamic viscosity, $\delta = \O(\sqrt{\mu})$ is the boundary layer thickness, $t_0$ is the characteristic time, $a_0$ is the characteristic wave amplitude and $\ell$ is the wavelength.

We assume that the flow is governed by the incompressible Navier-Stokes equations:
\begin{eqnarray*}
 	\div\u &=& 0, \\
 	\pd{\u}{t} + \u\cdot\grad\u + \frac{1}{\rho}\grad p &=& \g + \frac{1}{\rho}\div\ttau,
\end{eqnarray*}
where $\ttau$ is the viscous stress tensor
\begin{equation*}
  \tau_{ij} = 2\mu\eps_{ij}, \quad \eps_{ij} = \frac12\Bigl(\pd{u_i}{x_j} + \pd{u_j}{x_i}\Bigr) \;.
\end{equation*}

We multiply the second equation by $\u$ and integrate over the domain $\Omega$ with boundary $\partial\Omega$ to get the following energy balance equation:
\begin{multline*}
  \frac12\int\limits_{\Omega} \pd{}{t}\bigl(\rho|\u|^2\bigr)\; d\Omega + 
  \frac12\int\limits_{\partial\Omega} \rho|\u|^2 \u\cdot\n\; d\sigma = \\ =
  \int\limits_{\partial\Omega} \bigl(-p\I + \ttau\bigr)\n\cdot\u\; d\sigma +
  \int\limits_{\Omega}\rho\g\cdot\u\; d\Omega -\underbrace{\frac{1}{2\mu}\int\limits_{\Omega} \ttau : \ttau\; d\Omega}_{\T}\; .
\end{multline*}
In this identity each term has a precise physical meaning. The left-hand side is the total rate of energy change in $\Omega$. The second term is the flux of energy across the boundary. On the right-hand side, the first integral represents the rate of work by surface stresses acting on the boundary. The second integral is the rate of work done by the gravity force throughout the volume, and the third integral $\T$ is the rate of viscous dissipation. We focus our attention on the last term $\T$. We estimate the order of magnitude of the rate of dissipation in various regions of the fluid.

We start by the interior region $R_i$. Outside the boundary layers, it is reasonable to expect that the rate of strain is dominated by the irrotational part of the velocity whose scale is $\frac{a_0}{t_0}$ and the length scale is the wavelength $\ell$. The energy dissipation rate is then
\begin{equation*}
  \O\bigl(\T_{R_i}\bigr) \sim \frac{1}{\mu} \Bigl(\mu\frac{a}{t_0\ell}\Bigr)^2 \cdot \ell^3 = \mu\Bigl(\frac{a}{t_0}\Bigr)^2\ell \sim \O(\mu) \;.
\end{equation*}

Inside the bottom boundary layer the normal gradient of the solenoidal part of $\u$ dominates the strain rate, so that
\begin{equation*}
  \O\bigl(\T_{R_b}\bigr) \sim \frac{1}{\mu} \Bigl(\mu\frac{a}{t_0\delta}\Bigr)^2 \cdot \delta\ell^2 = \frac{\mu}{\delta}\Bigl(\frac{a\ell}{t_0}\Bigr)^2 \sim \O(\mu^{\frac12}) \; .
\end{equation*}

A free surface boundary layer also exists. Its importance depends on the free surface conditions. Consider first the classical case of a clean surface. The stress is mainly controlled by the potential velocity field which is of the same order as in the main body of the fluid. Because of the small volume $\O(\delta\ell^2)$ the rate of dissipation in the free surface boundary layer is only
\begin{equation*}
  \O\bigl(\T_{R_f}\bigr) \sim \frac{1}{\mu} \Bigl(\mu\frac{a}{t_0\ell}\Bigr)^2 \cdot \delta\ell^2
  = \mu\delta\Bigl(\frac{a}{t_0}\Bigr)^2 \sim \O(\mu^{\frac32})\;.
\end{equation*}
From the physical point of view it is weaker, since only the zero shear stress condition on the free surface is required.

Another extreme case is when the free surface is heavily contaminated, for example, by oil slicks. The stress in the free surface boundary layer can then be as great as in the boundary layer near a solid wall. In the present study we do not treat such extreme situations and the surface contamination is assumed to be absent.

The previous scalings suggest the following diagram which represents the hierarchy of dissipative terms:
\begin{equation*}
  \underbrace{\O\bigl(\mu^{\frac12}\bigr)}_{R_b} \hookrightarrow
  \underbrace{\O(\mu)}_{R_i} \hookrightarrow
  \underbrace{\O\bigl(\mu^\frac32\bigr)}_{R_f} \hookrightarrow \ldots
\end{equation*}
It is clear that the largest energy dissipation takes place inside the bottom boundary layer. We take into account only two first phenomena from this diagram. Consequently, all dissipative terms of order $\O(\mu^{\frac32})$ and higher will be neglected.

\begin{rem}
In laboratory experiments, surface waves are confined by side walls as well. The rate of attenuation due to the side-wall boundary layers was computed in \cite{Mei1973}. For simplicity, in the present study we consider an unbounded in horizontal coordinates domain (see \figurename~\ref{fig:anatomy}).
\end{rem}


\section{Derivation}\label{sec:derivation}

Consider the linearized 3D incompressible NS equations describing free-surface flows in a fluid layer of uniform depth $h$:
\begin{equation}\label{eq:NS}
  \pd{\v}{t} = -\frac{1}{\rho}\grad p + \nu\Delta\v + \g, \qquad \div\v = 0,
\end{equation}
with $\v$ the velocity vector, $p$ the pressure, $\rho$ the fluid density and $\g$ the acceleration due to gravity. We represent $\v=(u,v,w)$ in the form of the Helmholtz--Leray decomposition:
\begin{equation}\label{eq:helmleray}
  \v = \grad\phi + \grad\times\vpsi, \qquad \vpsi = (\psi_1, \psi_2, \psi_3).
\end{equation}
After substitution of the decomposition (\ref{eq:helmleray}) into (\ref{eq:NS}), one notices that the equations are verified provided that the functions $\phi$ and $\vpsi$ satisfy the following equations:
\begin{equation*}
  \Delta\phi = 0, \qquad
  \phi_{t} + \frac{p-p_0}{\rho} + gz = 0, \qquad
  \pd{\vpsi}{t} = \nu\Delta\vpsi.
\end{equation*}

Next we discuss the boundary conditions. We assume that the velocity field satisfies the conventional no-slip condition at the bottom $\left.\v\right|_{z=-h} = \0$, while at the free surface we have the usual kinematic condition, which can be stated as 
$$
	\eta_t + \v\cdot\grad\eta = w.
$$
After linearization it becomes simply $\eta_t = w$. 

Dynamic condition states that the forces must be equal on both sides of the free surface:
\begin{equation*}
	[\ssigma\cdot\n] \equiv -(p-p_0)\vec n + \ttau \cdot \vec n = 0 \quad \mbox{at} \;\; z=\eta(x,t) \,,
\end{equation*}
where $\ssigma$ is the stress tensor, $[f]$ denotes the jump of a function $f$ across the free surface, $\vec n$ is the normal to the free surface and $\ttau$ the viscous part of the stress tensor $\ssigma$.

The basic idea consists in expressing the vortical part of the velocity field $\grad\times\vpsi$ in terms of the velocity potential $\phi$ and the free surface elevation $\eta$ using differential or pseudodifferential operators. In this section we just show final results while the details of computation can be found in \cite{DutykhDias2007, Dutykh2007a}.

Let us begin by the free-surface kinematic condition 
$$
	\pd{\eta}{t} = w \equiv \pd{\phi}{z} + \pd{\psi_2}{x} - \pd{\psi_1}{y},
	\quad z=0.
$$

Using the absence of tangential stresses on the free surface, one can replace the rotational part in the kinematic boundary condition:
\begin{equation*}
  \eta_{t} = \phi_{z} + 2\nu\Delta\eta.
\end{equation*}

In order to account for the presence of viscous stresses, we have to modify the dynamic free-surface condition as well. This is done using the balance of normal stresses at the free surface:
\begin{equation*}
  \sigma_{zz} = 0 \mbox{ at } z=0 \Rightarrow
  p-p_0 = 2\rho\nu\pd{w}{z} \equiv 2\rho\nu\Bigl(
  \pd{^2\phi}{z^2} + \pd{^2\psi_2}{x\partial z} - \pd{^2\psi_1}{y\partial z}\Bigr).
\end{equation*}
In \cite{DutykhDias2007} it is shown that 
$\pd{^2\psi_2}{x\partial z} - \pd{^2\psi_1}{y\partial z} = \O(\nu^\frac12)$, so Bernoulli's equation becomes 
\begin{equation}\label{berdiss}
  \phi_{t} + g\eta + 2\nu\phi_{zz} + \O(\nu^\frac32) = 0.
\end{equation}
Since we only consider weak dissipation ($\nu\sim 10^{-6} - 10^{-3}$ m$^2$/s), we neglect terms of order $o(\nu)$.

The second step in our derivation consists in introducing a boundary layer correction at the bottom. In the present study we assume the boundary layer to be laminar. The question of turbulent boundary layer was investigated in \cite{Liu2006a}. Hence, the bottom boundary condition becomes
\begin{equation}\label{fond_dissip}
  \left.\pd{\phi}{z}\right|_{z=-h} = \sqrt{\frac{\nu}{\pi}}\int\limits_0^t\frac{\left.\nabla^2_{\x}\phi_0\right|_{z=-h}}{\sqrt{t-\tau}}\; d\tau = -\sqrt{\frac{\nu}{\pi}}\int\limits_0^t\frac{\left.\phi_{0zz}\right|_{z=-h}}{\sqrt{t-\tau}}\; d\tau.
\end{equation}
One recognizes on the right-hand side a half-order integral operator. The last equation shows that the effect of the diffusion process in the boundary layer is not instantaneous. The result is cumulative and it is weighted by $(t-\tau)^{-\frac12}$ in favour of the present time.

Summarizing the developments made above and generalizing our equations by including nonlinear terms, we obtain a set of viscous potential free-surface flow equations:
\begin{eqnarray} \label{full1}
  \Delta\phi &=& 0, \qquad\qquad (\x,z) \in \Omega = \R^2\times[-h,\eta] \\
  \eta_{t} + \grad\eta\cdot\grad\phi &=& \phi_{z} + 2\nu\Delta\eta, \qquad z=\eta \\
  \phi_{t} + \frac12\abs{\grad\phi}^2 + g\eta &=& -2\nu\phi_{zz}, \qquad\qquad z=\eta \\ \label{full2}
  \phi_{z} &=& -\sqrt{\frac{\nu}{\pi}}
  \int\limits_0^t\frac{\phi_{zz}}{\sqrt{t-\tau}}\; d\tau,
  \qquad z = -h.
\end{eqnarray}
At the present stage, the addition of nonlinear terms is rather a conjecture. However, a recent study by Liu et al. \cite{Liu2007} suggests that this conjecture is rather true. The authors investigated the importance of nonlinearity in the bottom boundary layer for a solitary wave solution. They came to the conclusion that ``the nonlinear effects are not very significant''.

Using this weakly damped potential flow formulation and described in previous works procedure \cite[Section 4]{Dutykh2007} of Boussinesq equations derivation, one can derive the following system of equations with horizontal velocity $\u_\theta$ defined at the depth $z_\theta=-\theta h$, $ 0 \leq \theta \leq 1$: 
\begin{equation}\label{bouss1}
  \eta_t + \div\left((h+\eta)\u_\theta\right) + h^3\left(\frac{\theta^2}{2}-\theta+\frac13\right)
  \nabla^2(\div\u_\theta) = 2\nu\Delta\eta + 
  \sqrt{\frac{\nu}{\pi}}\int\limits_0^t\frac{\div\u_\theta}{\sqrt{t-\tau}}\; d\tau,
\end{equation}
\begin{equation}\label{bouss2}
  \u_{\theta t} + \frac12\nabla|\u_\theta|^2 + g\nabla\eta - h^2\theta\left(1-\frac{\theta}{2}\right)\nabla(\div\u_{\theta t})
  = 2\nu\Delta\u_\theta.
\end{equation}
For simplicity, in this study we present governing equations only on the flat bottom, but generalization can be done for general bathymetry. Khabakhpashev \cite{Khabakhpashev1987}, Liu \& Orfila \cite{Liu2004} also derived a similar set of Boussinesq equations in terms of depth-averaged velocity. Our equations (\ref{bouss1}) -- (\ref{bouss2}) have local dissipative terms in addition and they are formulated in terms of the velocity variable defined at arbitrary water level that is beneficial for dispersion relation properties.


\subsection{Total energy decay in visco-potential flow}

Let us consider a fluid layer infinite in horizontal coordinates $\x = (x,y)$, bounded below by the flat bottom $z= -h$ and above by the free surface. Total energy of water waves is given by the following formula:
\begin{equation*}
  \E = \iint\limits_{\R^2}\int\limits_{-h}^{\eta} \frac12|\grad\phi|^2 dz d\x + \frac{g}{2}\iint\limits_{\R^2}\eta^2 d\x.
\end{equation*}
We would like to mention here that Zakharov showed \cite{Zakharov1968} this expression to be the Hamiltonian for classical water wave problem with suitable choice of canonical variables: $\eta$ and $\psi := \phi (\x, z = \eta, t)$.

In this section we are interested in the evolution of the total energy $\E$ with time. This question is investigated by computing the derivative $\od{\E}{t}$ with respect to time $t$. Obviously, when one considers the classical potential free surface flow formulation \cite{Lamb1932}, we have $\od{\E}{t} \equiv 0$, since no dissipation is introduced into the governing equations. We performed the computation of $\od{\E}{t}$ in the framework of the visco-potential formulation and give here the final result:

\begin{equation*}
  \od{\E}{t} = \sqrt{\frac{\nu}{\pi}}\iint\limits_{\R^2}\left.\phi_t\right|_{z=-h}\int\limits_0^t
  \frac{\left.\phi_{zz}\right|_{z=-h}}{\sqrt{t-\tau}}\;d\tau d\x + 
  2\nu\iint\limits_{\R^2} \partial_t(\left.\grad\phi\right|_{z=\eta}\cdot\grad\eta) \; d\x.
\end{equation*}

In this identity, the first term on the right hand side comes from the boundary layer and is predominant in the energy decay since its magnitude scales with $\O(\sqrt{\nu})$. The second term has its origins in free surface boundary conditions. Its magnitude is $\O(\nu)$, thus it has less important impact on the energy balance. This topic will be investigated further in future studies.


\subsection{Dissipative KdV equation}\label{sec:viscouskdv}

In this section we derive a viscous Korteweg-de Vries (KdV) equation from just obtained Boussinesq equations (\ref{bouss1}), (\ref{bouss2}). Since KdV-type equations model only unidirectional wave propagation, our attention is naturally restricted to 1D case. In order to perform asymptotic computations, all the equations have to be switched to nondimensional variables as it is explained in \cite[Section 2]{Dutykh2007}. We find the velocity variable $u$ in this form:
\begin{equation*}
  u_\theta = \eta + \eps P + \mu^2 Q + \ldots
\end{equation*}
where $\eps$, $\mu$ are nonlinearity and dispersion parameters respectively (see \cite{Dutykh2007} for their definition), $P$ and $Q$ are unknown at the present moment. Using the methods similar to those used in \cite[Section 6.1]{Dutykh2007}, one can easily show that
\begin{equation*}
 P = -\frac14\eta^2, \quad Q = \bigl(\theta-\frac16-\frac{\theta^2}{2}\bigr)\eta_{xx}.
\end{equation*}
This result immediately yields the following asymptotic representation of the velocity field
\begin{equation}\label{eq:velocityfield}
  u_\theta = \eta - \frac14\eps\eta^2 + \mu^2\bigl(\theta-\frac16-\frac{\theta^2}{2}\bigr)\eta_{xx} + \ldots
\end{equation}
Substituting the last formula (\ref{eq:velocityfield}) into equation (\ref{bouss1}) and switching again to dimensional variables, one obtains this viscous KdV-type equation:
\begin{equation}\label{eq:KdVviscous}
  \eta_t + \sqrt{\frac{g}{h}}\Bigl(
   (h+\frac32\eta)\eta_x + \frac16h^3\eta_{xxx} 
   - \sqrt{\frac{\nu}{\pi}}\int\limits_0^t\frac{\eta_x}{\sqrt{t-\tau}}\;d\tau\Bigr) = 2\nu\eta_{xx}.
\end{equation}
This equation will be used in Section \ref{sec:damping} to study the damping of linear progressive waves. Integral damping terms are reasonably well known in the context of KdV type equations. Various nonlocal corrections can be found, for example, in the following references \cite{Byatt-Smith1971a, Chester1968, Miles1976, Grimshaw2003}.

A similar nonlocal KdV equation was already derived in \cite{KM}. They used a different scaling in boundary layer which resulted in dissipative term nonlocal in space. Later, Matsuuchi \cite{Matsuuchi1976} performed a comparison of numerical computations with their model equation against laboratory data. They showed that their model does not reproduce well the phase shift:
\begin{quote}
  \dots it may be concluded that our modified K-dV equation can describe the observed wave behaviours except the fact that the phase shift obtained by the calculations is not confirmed by their experiments.
\end{quote}
Excellent performance of the model (\ref{eq:KdVviscous}) with respect to experiments was shown in \cite{Liu2006}.


\section{Dispersion relation of complete and Boussinesq nonlocal equations}\label{sec:disprel}

Interesting information about the governing equations can be obtained from the linear dispersion relation analysis. In this section we are going to analyse the new set of equations (\ref{full1})--(\ref{full2}) for the complete water wave problem and the corresponding long wave asymptotic limit (\ref{bouss1}), (\ref{bouss2}).

To simplify the computations, we consider the two-dimensional problem. The generalization to higher dimensions is straightforward and is performed by replacing the wavenumber $k$ by its modulus $|\vec{k}|$ in vectorial case. Traditionally the governing equations are linearized and the bottom is assumed to be flat. The last hypothesis is made throughout this study. After all these simplifications the new set of equations becomes
\begin{eqnarray}
  \phi_{xx} + \phi_{zz} = 0, \qquad (x,z)&\in & \R\times[-h,0], \label{eq:cont} \\
  \eta_t = \phi_z + 2\nu\eta_{xx}, \qquad z &=& 0, \label{eq:freesurf} \\
  \phi_t + g\eta + 2\nu\phi_{zz} = 0, \qquad z &=& 0, \\
  \phi_z + \sqrt{\frac{\nu}{\pi}}\int\limits_0^t
  \frac{\phi_{zz}}{\sqrt{t-\tau}}\; d\tau = 0, \qquad z&=&-h. \label{eq:bottom}
\end{eqnarray}
The next classical step consists in finding solutions of the special form
\begin{equation}\label{eq:specialform}
  \phi(x,z,t) = \varphi(z) e^{i(kx-\omega t)}, \qquad
  \eta(x,t) = \eta_0 e^{i(kx-\omega t)}.
\end{equation}
From continuity equation (\ref{eq:cont}) we can determine the structure of the function $\varphi(z)$:
\begin{equation*}
  \varphi(z) = C_1 e^{kz} + C_2 e^{-kz}.
\end{equation*}
Altogether we have three unknown constants\footnote{Since the present problem is linear, we have effectively only two degrees of freedom but it is not important for our purposes.} $\vec C = (C_1, C_2, \eta_0)$ and three boundary conditions (\ref{eq:freesurf})--(\ref{eq:bottom}) which can be viewed as a linear system with respect to $\vec C$:
\begin{equation}\label{eq:linsystem}
  \M\vec C = \vec 0.
\end{equation}
The matrix $\M$ has the following elements
\begin{equation*}
  \M = \begin{pmatrix}
        k & -k & i\omega - 2\nu k^2 \\
        i\omega - 2\nu k^2 & i\omega - 2\nu k^2 & -g \\
        e^{-kh}\bigl(1+kF(t,\omega)\bigr) & e^{kh}\bigl(-1+kF(t,\omega)\bigr) & 0 \\
       \end{pmatrix}
\end{equation*}
where the function $F(t,\omega)$ is defined in the following way:
\begin{equation}\label{eq:Ftomega}
 F(t,\omega) := \sqrt{\frac{\nu}{\pi}}\int\limits_0^t
 \frac{e^{i\omega(\tau)(t-\tau)}}{\sqrt{t-\tau}}\;d\tau.
\end{equation}
In order to have nontrivial solutions of (\ref{eq:cont})--(\ref{eq:bottom}), the determinant of the system (\ref{eq:linsystem}) has to be equal to zero $\det\M = 0$. It gives us a relation between $\omega$ and wavenumber $k$. This relation is called the linear dispersion relation:
\begin{multline}\label{eq:domegak}
  D(\omega,k) := (i\omega-2\nu k^2)^2 + gk\tanh(kh) \\
  - k F(t,\omega)\bigl((i\omega-2\nu k^2)^2\tanh(kh) + gk\bigr) \equiv 0.
\end{multline}

A similar procedure can be followed for Boussinesq equations (\ref{bouss1}), (\ref{bouss2}). We do not give here the details of the computations but only the final result:
\begin{multline}\label{eq:dbomegak}
  D_{b}(\omega,k) := (i\omega - 2\nu k^2)^2 + b(kh)^2 i\omega(i\omega-2\nu k^2)
  \\ + gh k^2(1 - a(kh)^2) - gk^2F(t,\omega)\equiv 0,
\end{multline}
where we introduced the following notation: $a := \frac{\theta^2}{2} - \theta + \frac13$, $b := \theta\bigl(1-\frac{\theta}{2}\bigr)$.

Unfortunately, the relations $D(\omega,k)\equiv 0$ and $D_b(\omega,k)\equiv 0$ cannot be solved analytically to give an explicit dependence of $\omega$ on $k$ as for the classical water wave problem. Consequently, we apply a quadrature formula to discretize the nonlocal term $F(t_n,\omega)$, where $t_n$ denotes the discrete time variable. The resulting algebraic equation with respect to $\omega(t_n;k)$ is solved analytically.

\begin{rem}\label{rem:omega}
Contrary to the classical water wave problem and, by consequence, standard Boussinesq equations (their dispersion relation can be found in \cite[Section 3.2]{Dutykh2007}, for example) where the dispersion relation does not depend on time
\begin{equation}\label{eq:classical}
  \omega^2 - gk\tanh(kh) \equiv 0,
\end{equation}
here we have additionally the dependence of $\omega(k; t)$ on time $t$ as a parameter. It is a consequence of the presence of the nonlocal term in time in the bottom boundary condition (\ref{full2}). Physically it means that the boundary layer ``remembers'' the flow history.
\end{rem}

\begin{rem}\label{rem:time}
There is one subtle point in the derivation presented above. In fact, all computations were performed as if the frequency $\omega$ were independent from time. Our final result shows that time $t$ appears explicitly in the dispersion relations (\ref{eq:domegak}), (\ref{eq:dbomegak}). Developments made above make sense under the assumption of slow variation of $\omega$ with time $t$. This statement can be written in mathematical form $\pd{\omega}{t} \ll 1$. It is rather a conjecture here and will be examined in future studies. We had to make this assumption in order to avoid complicated integro-differential equations and, consequently, simplify the analysis.
\end{rem}


\subsection{Analytical limit for infinite time}

In the previous section we showed that the dispersion relation of our visco-potential formulation is time dependent. It is natural to ask what happens when time evolves. Here we compute the limiting state of the dispersion curves (\ref{eq:domegak}), (\ref{eq:dbomegak}) as $t\to+\infty$. Namely, we will take this limit in equations (\ref{eq:domegak}), (\ref{eq:dbomegak}) assuming, of course, that it exists \begin{equation*}
  \exists\;\omega_\infty(k) := \lim_{t\to+\infty}\omega_t(k).
\end{equation*}

Time $t$ comes in dispersion relations through the argument of the function $F(t,\omega)$ defined in (\ref{eq:Ftomega}). Its limit can be easily computed to give
\begin{equation*}
  \lim_{t\to+\infty} F(t,\omega) = \sqrt{\frac{\nu}{\pi\omega_\infty}}\int\limits_0^{+\infty} \frac{e^{ip}}{\sqrt{p}}\; dp = 
  \sqrt{\frac{\nu}{\omega_\infty}} e^{i\frac{\pi}{4}}.
\end{equation*}

Now we are ready to write down the final results:
\begin{multline*}
  D(\omega_\infty, k) := (i\omega_\infty - 2\nu k^2)^2 + gk\tanh(kh) \\ - \sqrt{\frac{\nu}{\omega_\infty}}k e^{i\frac\pi4}
  \bigl((i\omega_\infty - 2\nu k^2)^2\tanh(kh) + gk\bigr) \equiv 0,
\end{multline*}

\begin{multline*}
  D_{b}(\omega_\infty, k) := (i\omega_\infty - 2\nu k^2)^2 + b(kh)^2 i\omega_\infty(i\omega_\infty-2\nu k^2)
  \\ + gh k^2(1 - a(kh)^2) - gk^2\sqrt{\frac{\nu}{\omega_\infty}}e^{i\frac\pi4} \equiv 0.
\end{multline*}

In order to solve numerically nonlinear equation $D(\omega_\infty, k) = 0$ (or $D_b(\omega_\infty, k) = 0$ when one is interested in Boussinesq equations) with respect to $\omega_\infty$, we apply a Newton-type method. The iterations converge very quickly since we use analytical expressions for the Jacobian $\od{D}{\omega_\infty}$ ($\od{D_b}{\omega_\infty}$, correspondingly). Derivative computation is straightforward. Limiting dispersion curves are plotted (see \figurename~\ref{fig:infty}) and discussed in the next section.


\subsection{Discussion}

\begin{figure}
	\centering
		\includegraphics[width=0.70\textwidth]{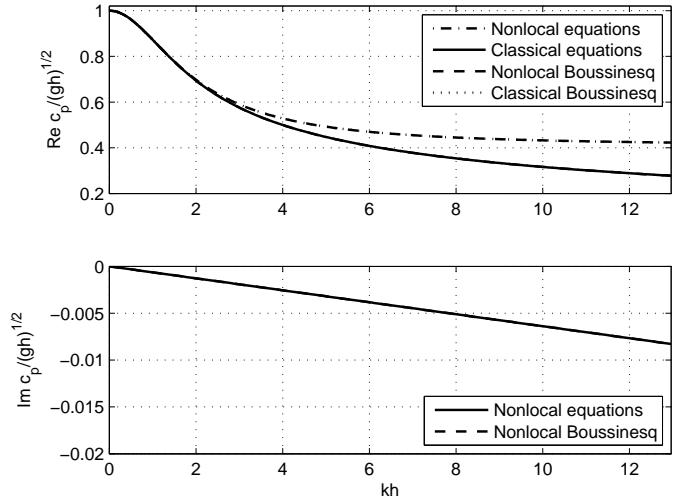}
	\caption[Real and imaginary part of dispersion curve at $t=0$.]{Real and imaginary part of dispersion curve at $t=0$. At the beginning,	the nonlocal term has no effect. Thus, the real parts of the classical water wave problem and new set of equations are exactly superimposed on this figure. The imaginary part represents only local dissipation at this stage.}
	\label{fig:DispRelT=0}
\end{figure}

\begin{figure}
	\centering
		\includegraphics[width=0.70\textwidth]{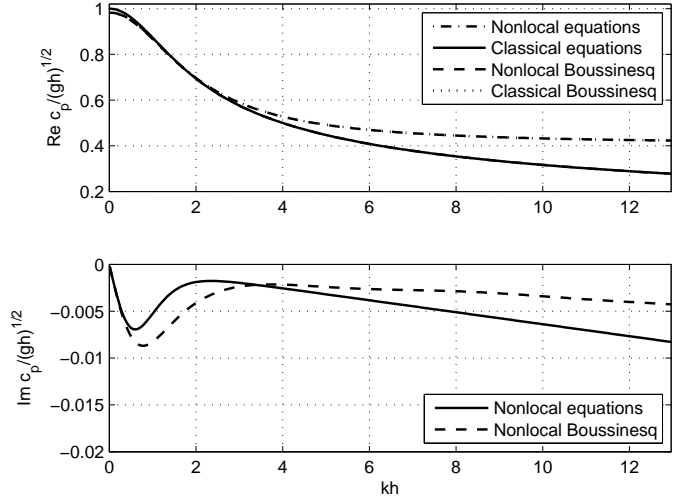}
	\caption[Phase velocity at $t=1$.]{Phase velocity at $t=1$. Boundary layer effects start to be visible: the real part of the velocity slightly drops down and the straight lines of the imaginary part are deformed by the nonlocal term. Within graphical accuracy, the classical and their nonlocal counterparts are superimposed.}
	\label{fig:DispRelT=1}
\end{figure}

\begin{figure}
	\centering
		\includegraphics[width=0.70\textwidth]{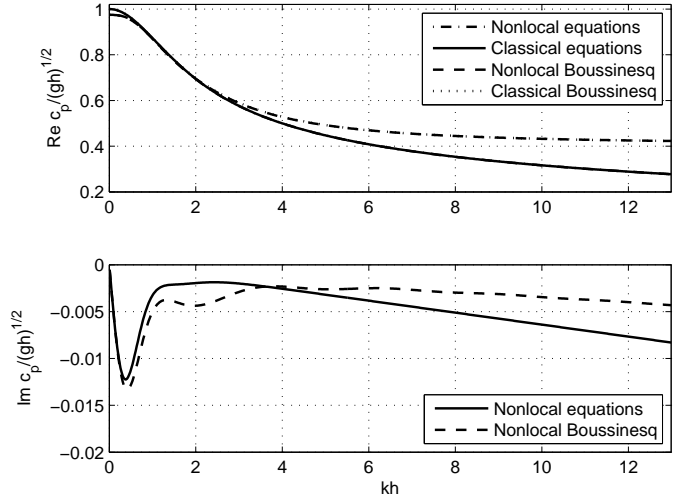}
	\caption[Phase velocity at $t=2$.]{Phase velocity at $t=2$. Nonlocal term slows down long waves since the real part of the phase velocity decreases.}
	\label{fig:DispRelT=2}
\end{figure}

\begin{figure}
	\centering
		\includegraphics[width=0.70\textwidth]{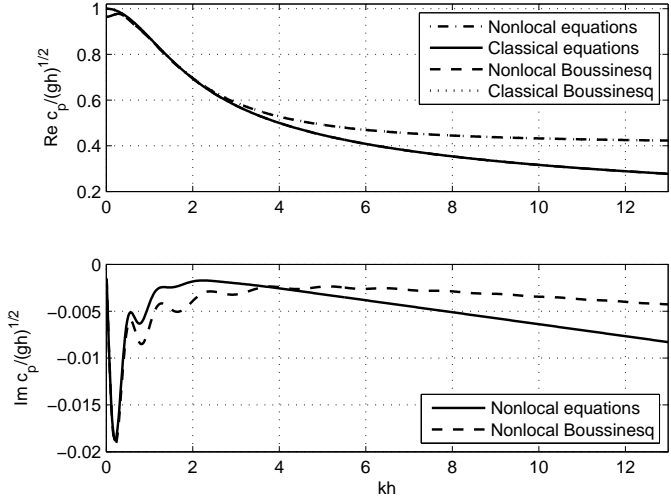}
	\caption[Phase velocity at $t=4$.]{Phase velocity at $t=4$. On \figurename~\ref{fig:DispRelT=4Zoom} we plot a zoom on long and moderate waves.}
	\label{fig:DispRelT=4}
\end{figure}

\begin{figure}
	\centering
	\subfigure[Zoom on long waves. Classical and corresponding Boussinesq equations are almost superimposed in this region.]%
	{\includegraphics[width=0.48\textwidth]{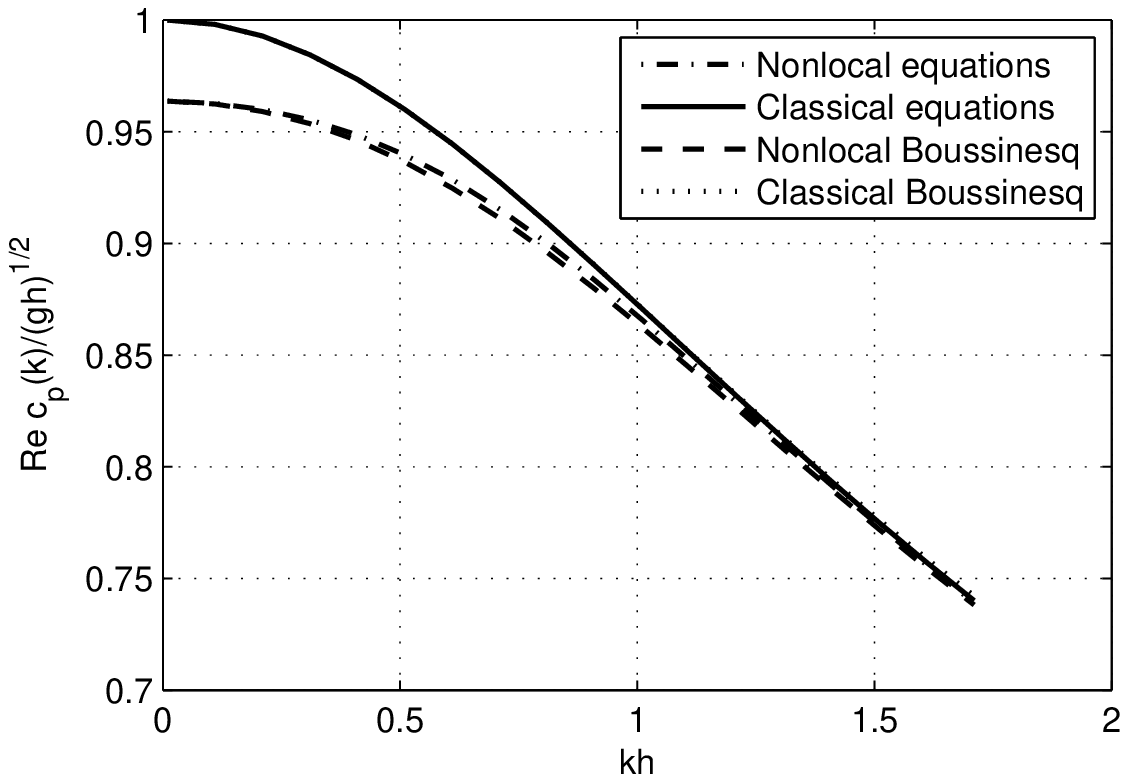}} \quad
	\subfigure[Zoom on moderate wavelengths. In this region classical and nonlocal complete equations are almost superimposed. In the same time one can notice a little difference in Boussinesq models.]%
	{\includegraphics[width=0.48\textwidth]{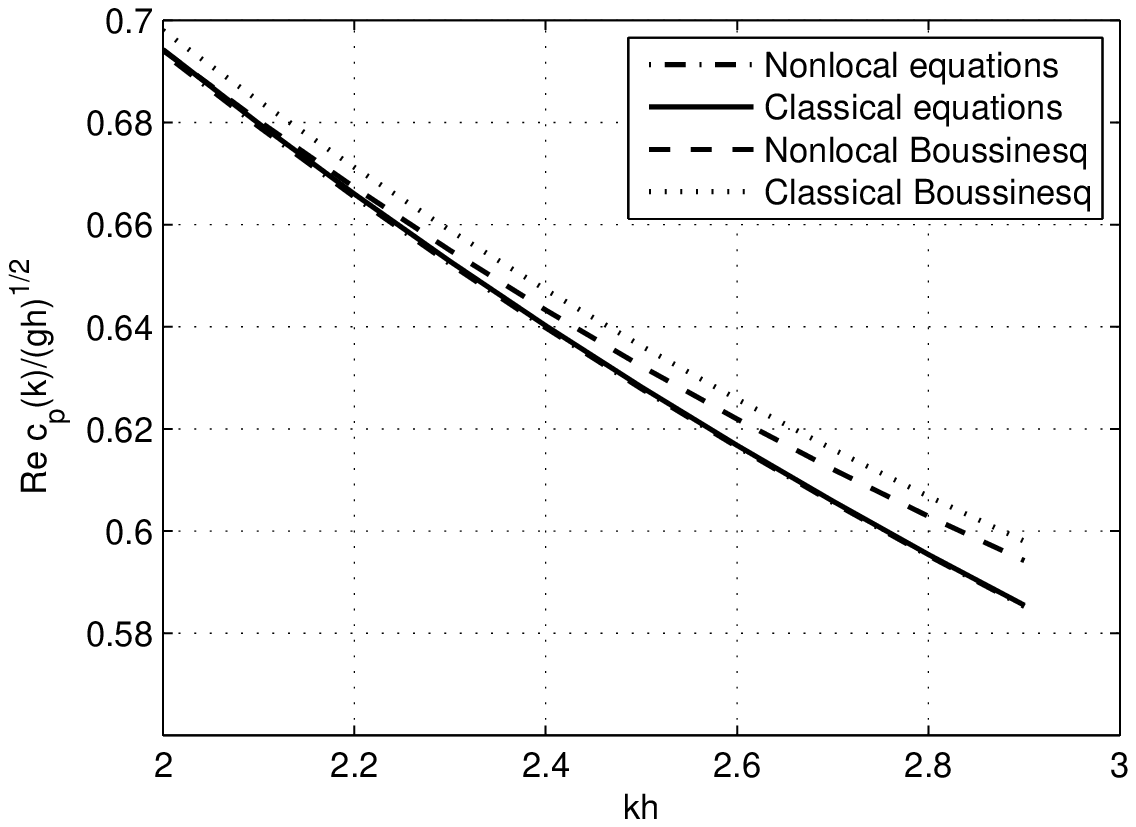}}
	\caption{Real part of the phase velocity at $t=4$.}
	\label{fig:DispRelT=4Zoom}
\end{figure}

Numerical snapshots of the nonclassical dispersion relation\footnote{To be precise, we plot real and imaginary parts of the dimensionless phase velocity which is defined as $c_p(k) := \frac{1}{\sqrt{gh}}\frac{\omega(k)}{k}$.} at different times for complete and Boussinesq equations are given on Figures (\ref{fig:DispRelT=0})--(\ref{fig:DispRelT=4Zoom}). The value of the eddy viscosity $\nu$ is taken from Table \ref{tab:parametersDamping} and we consider a one meter depth fluid layer ($h = 1$ m). We will try to make several comments on the obtained results.

\begin{rem}
Recall that these snapshots were obtained under the assumption that $\omega$ is slowly varying in time. The validity of this approximation is examined and discussed in \citep{Dutykh2008b}.
\end{rem}

Just at the beginning (when $t=0$), there is no effect of the nonlocal term. This is why on \figurename~\ref{fig:DispRelT=0} new and classical\footnote{In this section expression ``classical'' refers to complete water wave problem or its dispersion relation correspondingly.} curves are superimposed. With no surprise, the phase velocity of Boussinesq  equations represents well only long waves limit (let us say up to $kh\approx 2$). When time evolves, we can see that the main effect of nonlocal term consists in slowing down long waves (see Figures \ref{fig:DispRelT=1}--\ref{fig:DispRelT=4}). Namely, in the vicinity of $kh = 0$ the real part of the phase velocity is slightly smaller with respect to the classical formulation. From physical point of view this situation is comprehensible since only long waves ``feel'' the bottom and, by consequence, are affected by bottom boundary layer. On the other hand, the imaginary part of the phase velocity is responsible for the wave amplitude attenuation. The minimum of $\Im c_p(k)$ in the region of long waves indicates that there is a ``preferred'' wavelength which is attenuated the most. In the range of short waves the imaginary part is monotonically decreasing. In practice it means that high-frequency components are damped by the model. This property can be advantageous in numerics, for example. On \figurename~\ref{fig:DispRelT=4Zoom} we depicted the real part of $c_p(k)$ with zoom made on long and moderate waves. The reader can see that nonlocal full and Boussinesq equations have similar behaviour in the vicinity of $kh = 0$.

\begin{figure}
	\centering
		\includegraphics[width=0.70\textwidth]{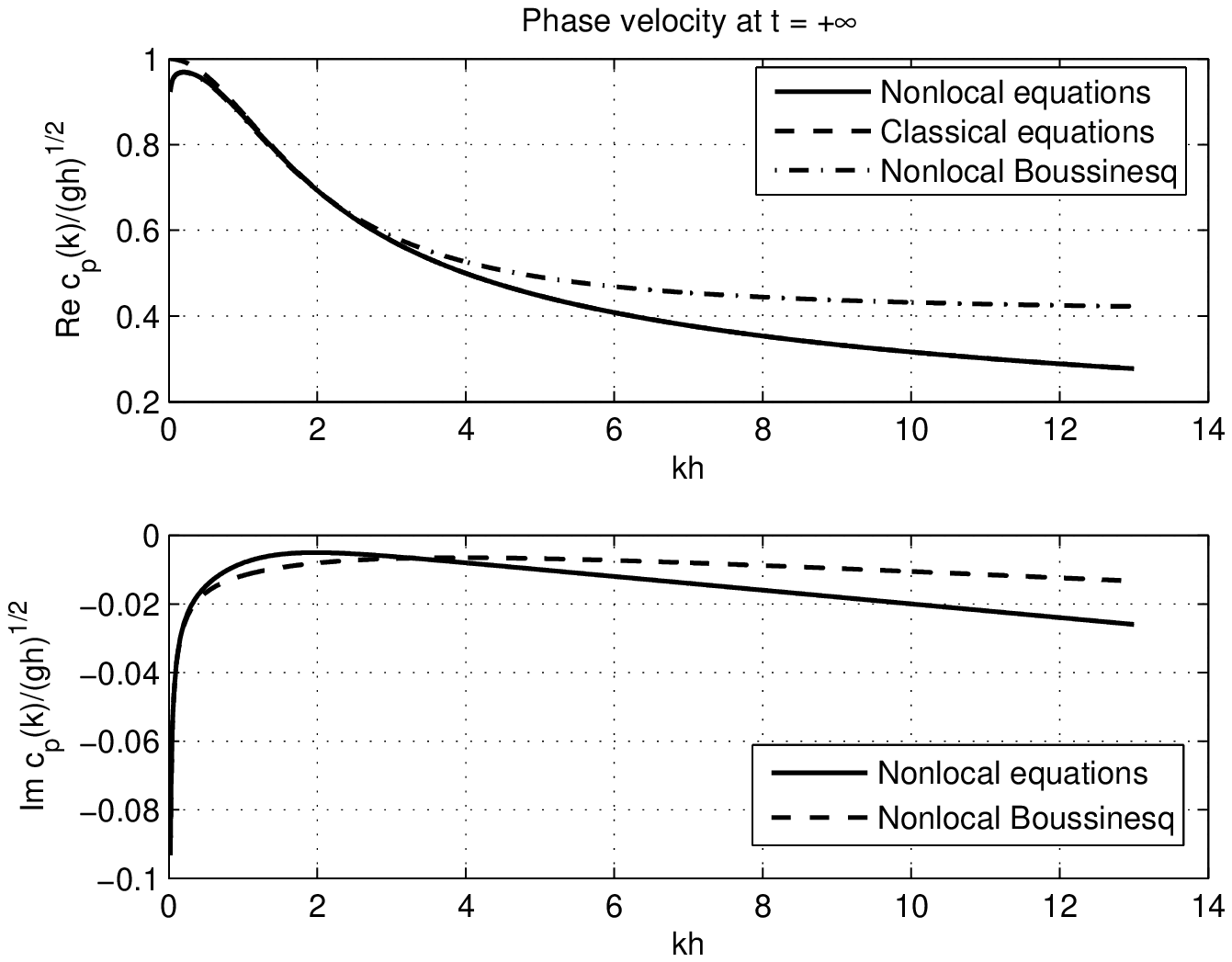}
	\caption{Limiting steady state of the phase velocity at $t\to+\infty$.}
	\label{fig:infty}
\end{figure}

Now let us discuss the limiting state of phase velocity curves as $t\to+\infty$. It is depicted on \figurename~\ref{fig:infty}. One can see singular behaviour in the vicinity of zero. This situation is completely normal since very long waves are highly affected by bottom boundary layer. 

We would like to comment more on the behaviour of curves on \figurename~\ref{fig:infty} since it is not easy to distinguish them with graphical resolution. We will concentrate on the upper image because everything is clear below with imaginary part. In the vicinity of $kh = 0$ we have the superposition of nonlocal models (complete set of equations (\ref{full1}) -- (\ref{full2}) and Boussinesq equations (\ref{bouss1}) -- (\ref{bouss2})). When we gradually move to short waves, we have the superposition of complete classical and nonlocal (\ref{full1}) -- (\ref{full2}) water wave problems. Meanwhile, Boussinesq system gives slightly different phase velocity for $kh \geq 3$. This is comprehensible since we cannot simplify considerably the problem and have uniformly good approximation everywhere. Various Boussinesq systems are designed to reproduce the behaviour of long waves. 

\section{Attenuation of linear progressive waves}\label{sec:damping}

In this Section we investigate the damping rate of linear progressive waves. Thus, the first step consists in linearizing dissipative KdV equation (\ref{eq:KdVviscous}) to obtain the following nonlocal Airy equation:
\begin{equation}\label{eq:linearKdV}
  \eta_t + \sqrt{\frac{g}{h}}\Bigl(h\eta_x + \frac16h^3\eta_{xxx} - \sqrt{\frac{\nu}{\pi}}\int\limits_0^t\frac{\eta_x}{\sqrt{t-\tau}}\;d\tau
  \Bigr) = 2\nu\eta_{xx}
\end{equation}
In other words, we can say that we restrict our attention only to small amplitude waves. Now we make the next assumption. We look for a particular form of the solutions:
\begin{equation}\label{eq:special}
  \eta(x,t) = \A(t)e^{ik\xi}, \quad \xi = x - \sqrt{gh} t.
\end{equation}
where $k$ is the wavenumber and $\A(t)$ is called the complex amplitude, since $|\eta(x,t)| = |\A(t)|$. Integro-differential equation governing the temporal evolution of $\A(t)$ can be easily derived by substituting the special representation (\ref{eq:special}) into linearized KdV equation (\ref{eq:linearKdV}):
\begin{equation}\label{eq:at}
  \od{\A}{t} - \frac{i}{6}\sqrt{\frac{g}{h}}(kh)^3 \A(t) + 2\nu k^2 \A(t) 
  -ik\sqrt{\frac{g\nu}{\pi h}}\int\limits_0^t\frac{\A(\tau)}{\sqrt{t-\tau}}\;d\tau = 0.
\end{equation}

In our applications we are rather interested in temporal evolution of the absolute value $|\A(t)|$. It is straightforward to derive the governing equation for the squared wave amplitude:
\begin{equation*}
  \od{|\A|^2}{t} + 4\nu k^2|\A(t)|^2 
  - ik\sqrt{\frac{g \nu}{\pi h}}\int\limits_0^t\frac{\bar\A(t)\A(\tau) - \A(t)\bar\A(\tau)}{\sqrt{t-\tau}}\;d\tau = 0.
\end{equation*}
If we denote by $\A_r(t)$ and $\A_i(t)$ real and imaginary parts of $\A(t)$ respectively, the last equation can be further simplified:
\begin{equation}\label{eq:squaredamplitude}
  \od{|\A|^2}{t} + 4\nu k^2|\A(t)|^2 
  + 2k\sqrt{\frac{g \nu}{\pi h}}\int\limits_0^t\frac{\A_r(t)\A_i(\tau) - \A_i(t)\A_r(\tau)}{\sqrt{t-\tau}}\;d\tau = 0.
\end{equation}
Just derived integro-differential equation represents a generalisation to the classical equation (\ref{eq:classical0}) by Boussinesq \cite{Boussinesq1895} and Lamb \cite{Lamb1932} for the periodic, linear wave amplitude evolution in a viscous fluid. We recall that novel integral term is a direct consequence of the bottom boundary layer modelling.

Unfortunately, equation (\ref{eq:squaredamplitude}) cannot be used directly for numerical computations since we need to know the following combination of real and imaginary parts $\A_r(t)\A_i(\tau) - \A_i(t)\A_r(\tau)$ for $\tau\in[0,t]$. It represents a new and non-classical aspect of the present theory.

Equation (\ref{eq:squaredamplitude}) allows us to discuss the relative importance of local and nonlocal dissipative terms for long waves. In fact, when we consider the deep-water approximation, only local dissipative terms are present in the governing equations \citep{Dias2007}. On the other hand, in shallow waters the integral term is predominant. It means that there is an intermediate depth where both dissipative terms have equal magnitude. This depth can be estimated when one switches to dimensionless form of the equation (\ref{eq:squaredamplitude}). Comparing the coefficients in front of dissipative terms gives the following transcendental equation for the ``critical'' depth $h^*$:
\begin{equation*}
  h^* = \frac{g}{4\pi\omega\nu k^2}
\end{equation*}
where $\omega$ is the characteristic wave frequency. 

In numerical computations it is advantageous to integrate exactly local terms in equation (\ref{eq:at}). It is done by making the following change of variables:
\begin{equation*}
  \A(t) = e^{-2\nu k^2 t}e^{\frac{i}{6}\sqrt{\frac{g}{h}}(kh)^3 t} \tilde\A(t).
\end{equation*}
One can easily show that new function $\tilde\A(t)$ satisfies the following equation:
\begin{equation*}
  \od{\tilde\A}{t} = ik\sqrt{\frac{g\nu}{\pi h}}
  \int\limits_0^t\frac{e^{2\nu k^2 (t-\tau)}e^{-\frac{i}{6}\sqrt{\frac{g}{h}}(kh)^3 
  (t-\tau)}}{\sqrt{t-\tau}}\tilde\A(\tau)\;d\tau
\end{equation*}

\begin{figure}
	\centering
		\includegraphics[width=0.70\textwidth]{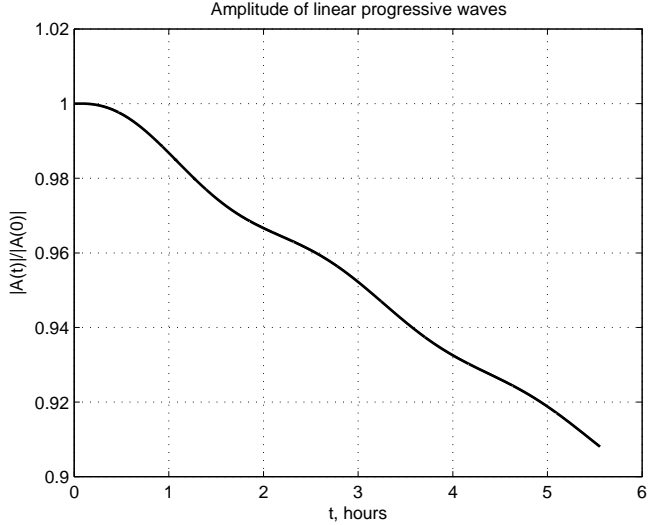}
	\caption{Amplitude of linear progressive waves as a function of time. Values of all parameters are given in Table \ref{tab:parametersDamping}.}
	\label{fig:fivehours}
\end{figure}

On \figurename~\ref{fig:fivehours} we plot a solution of integro-differential equation (\ref{eq:at}). All parameters related to this case are given in Table \ref{tab:parametersDamping}. These values were chosen to simulate a typical tsunami in Indian Ocean \cite{Dias2006}. We have to say that the wave amplitude damping is entirely due to the dissipation in boundary layer since local terms are unimportant for sufficiently long waves. It means that classical formula (\ref{eq:expdecay}) gives almost constant value $\alpha_0$ of the amplitude $\alpha(t)$ on the time scale of several hours, since the factor $2\nu k^2$ is of order $\approx 10^{-11}$ for parameters given in Table \ref{tab:parametersDamping}.

\begin{table}
	\begin{center}
		\begin{tabular}{ccc}
  		\hline\hline
  		\textit{parameter} & \textit{definition} & \textit{value} \\
      \hline
        $\nu$ & eddy viscosity & $10^{-3}$ $\frac{m^2}{s}$ \\
      \hline
        $g$ & gravity acceleration & $9.8$ $\frac{m}{s^2}$ \\
      \hline
       $h$ & water depth & 3600 $m$ \\
      \hline
       $\ell$ & wavelength & 50 $km$ \\
      \hline
       $k$ & wavenumber & $=\frac{2\pi}{\ell}$ $m^{-1}$ \\
      \hline\hline
		\end{tabular}
		\caption{Values of the parameters used in the numerical computations of the linear progressive waves amplitude. These values correspond to a typical Indian Ocean tsunami.}
		\label{tab:parametersDamping}
	\end{center}
\end{table}

\section{Solitary wave propagation}

In this section we would like to show the effect of nonlocal term on the solitary wave attenuation. For simplicity, we will consider wave propagation in a 1D channel. The question of the bottom shear stress effect on the solitary wave propagation was considered for the first time in \cite{Keulegan1948}.

For numerical integration of equations (\ref{bouss1}), (\ref{bouss2}) we use the same Fourier-type spectral method that was described in \cite[Section 5]{Dutykh2007}. Obviously this method has to be slightly adapted because of the presence of nonlocal in time term. We have to say that this term necessitates the storage of $\div\u^{(n)}$ at previous time steps. Hence, long time computations can be memory consuming.

The values of all dimensionless parameters are given in Table \ref{tab:parameters}. Dimensionless viscosity $\nu$ is related to other physical parameters in the following way:
\begin{equation*}
  \nu^2 = \frac{\bar\nu}{\ell\sqrt{gh}},
\end{equation*}
where $\bar\nu$ is kinematic viscosity, $\ell$ is the characteristic wavelength and $h$ is the typical depth.

\begin{rem}
  From numerical point of view, the integral term $\int\limits_0^t\frac{\phi_{zz}(\x,-h,\tau)}{\sqrt{t-\tau}}\; d\tau$ can pose some problems in the vicinity of the upper limit $\tau = t$. Probably the best way to deal with this problem is to separate the integral in two parts:
\begin{equation*}
  \int\limits_0^t\frac{\phi_{zz}(\x,-h,\tau)}{\sqrt{t-\tau}}\; d\tau =
  \int\limits_0^{t-\delta}\frac{\phi_{zz}(\x,-h,\tau)}{\sqrt{t-\tau}}\; d\tau + 
  \int\limits_{t-\delta}^t\frac{\phi_{zz}(\x,-h,\tau)}{\sqrt{t-\tau}}\; d\tau, \quad \delta > 0.
\end{equation*}
The first integral can be computed in a usual way without any special care. Then one applies to the second integral a special class of Gauss quadrature formulas with weighting function $(t-\tau)^{-\frac12}$. 

But there is another well-known trick that we describe here. This technique can be implemented in simpler way. We rewrite our integral in the following way
\begin{equation*}
  \int\limits_0^t\frac{\phi_{zz}(\x,-h,\tau)}{\sqrt{t-\tau}}\; d\tau = 
  \int\limits_0^t\frac{\phi_{zz}(\x,-h,t)}{\sqrt{t-\tau}}\; d\tau +
  \int\limits_0^t\frac{\phi_{zz}(\x,-h,\tau)-\phi_{zz}(\x,-h,t)}{\sqrt{t-\tau}}\; d\tau.
\end{equation*}
The first integral in the right hand side can be evaluated analytically while the second one does not contain any singularity under the assumption of differentiability of $\tau\mapsto\phi_{zz}(\x,-h,\tau)$ at $\tau = t$:
\begin{equation*}
  \int\limits_0^t\frac{\phi_{zz}(\x,-h,\tau)}{\sqrt{t-\tau}}\; d\tau = 
  2\sqrt{t}\phi_{zz}(\x,-h,t) + \int\limits_0^t\frac{\phi_{zz}(\x,-h,\tau)-\phi_{zz}(\x,-h,t)}{\sqrt{t-\tau}}\; d\tau.
\end{equation*}
\end{rem}

\begin{table}
	\begin{center}
		\begin{tabular}{ccc}
  		\hline\hline
  		\textit{parameter} & \textit{definition} & \textit{value} \\
      \hline
      	$\eps$ & nonlinearity & $0.02$ \\
      \hline
        $\mu$ & dispersion & $0.06$ \\
      \hline
        $\nu$ & eddy viscosity & $0.001$ \\
      \hline
        $c$ & soliton velocity & $1.02$ \\
      \hline
        $\theta$ & $z_\theta = -\theta h$ & $1-\sqrt{5}/5$ \\
      \hline
        $x_0$ & soliton center at $t=0$ & $-1.5$ \\
      \hline\hline
		\end{tabular}
		\caption{Values of the dimensionless parameters used in the numerical computations.}
		\label{tab:parameters}
	\end{center}
\end{table}

\begin{rem}\label{rem:eddy}
What is the value of $\nu$ to be taken in numerical simulations? There is surprisingly little published information of this subject. What is clear is that the molecular diffusion is too small to model true viscous damping and one should rather consider the eddy viscosity parameter. Some interesting information on this subject can be found in \cite{Tuck2007}:
\begin{quote}
  We have spent a considerable amount of time and effort seeking further published information on viscous effects on ship waves.
\end{quote}
This sentence confirms our apprehension. The authors of this work came to the following conclusion
\begin{quote}
   \dots we reiterate that a viscosity of about $\nu = 0.005$ $m^2/s$ gave reasonable agreement with longitudinal cut results (including apparent damping of transverse waves).
\end{quote}
In another famous paper \cite{Bona1981} one can find:
\begin{quote}
  \dots Such a decay rate leads to a value for $\mu$ in ($M^*$) of 0.014\dots
\end{quote}
In their work $\mu$ is the coefficient in front of the dissipative term in BBM equation:
\begin{equation*}
	  \eta_t + \eta_x + \frac32\eta\eta_x - \mu\eta_{xx} - \frac16\eta_{xxt} = 0.
\end{equation*}
It is important to underline that this equation is written in dimensionless variables. Thus, the value reported in that study has to be rescaled with respect to other physical parameters.

Liu and Orfilla \cite{Liu2004} report the value of eddy viscosity $\nu = 0.001$ $m^2/s$. 

If we summarize all the remarks made above, we can conclude that the value of the order $10^{-3}$ -- $10^{-2}$ $m^2/s$ should give reasonable results.
\end{rem}

\subsection{Approximate solitary wave solution}

In order to provide an initial condition for equations (\ref{bouss1}), (\ref{bouss2}), we are going to obtain an approximate solitary wave solution for nondissipative 1D version of these equations over the flat bottom:
\begin{eqnarray*}
  \eta_t + \bigl((1+\eps\eta)u\bigr)_x + \m\Bigl(\frac{\theta^2}{2}-\theta+\frac13\Bigr)u_{xxx} &=& 0, \\
  u_t + \eta_x + \frac{\eps}{2}(u^2)_x - \m\theta\Bigl(1-\frac{\theta}{2}\Bigr)u_{xxt} &=& 0.
\end{eqnarray*}

Then, we apply the same approach as in Section \ref{sec:viscouskdv} or in \cite[Section 6.1]{Dutykh2007}. We do not provide the computations here since they are simple and can be done without any difficulties. The final result is the following:
\begin{equation*}
  \eta(x,t) = \frac{2(c-1)}{\eps} \sech^2\Bigl(\frac{\sqrt{6(c-1)}}{2\mu}(x+x_0 - ct)\Bigr)
\end{equation*}
and the velocity is given by formula (\ref{eq:velocityfield}).
In the numerical results presented here, we use $\eta(x,0)$ and $u(x,0)$ as initial conditions.

\subsection{Discussion}

\begin{figure}
	\centering
		\includegraphics[width=0.70\textwidth]{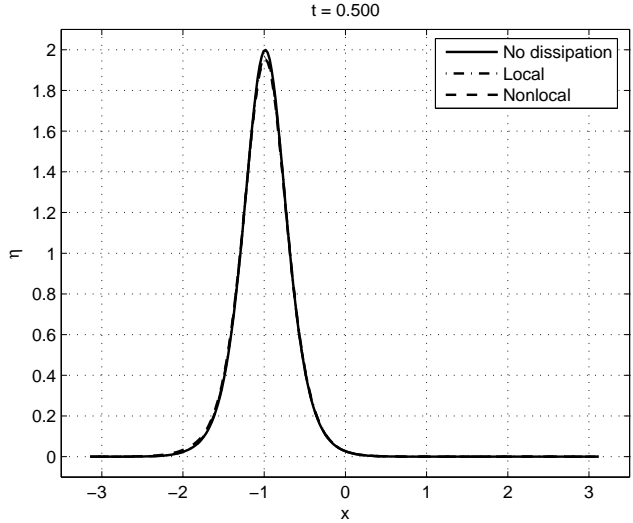}
	\caption[Snapshots of the free surface at $t=0.5$]{Comparison among two dissipative and nondissipative Boussinesq equations. Snapshots of the free surface at $t=0.5$}
	\label{fig:t0_5}
\end{figure}

\begin{figure}
	\centering
		\includegraphics[width=0.70\textwidth]{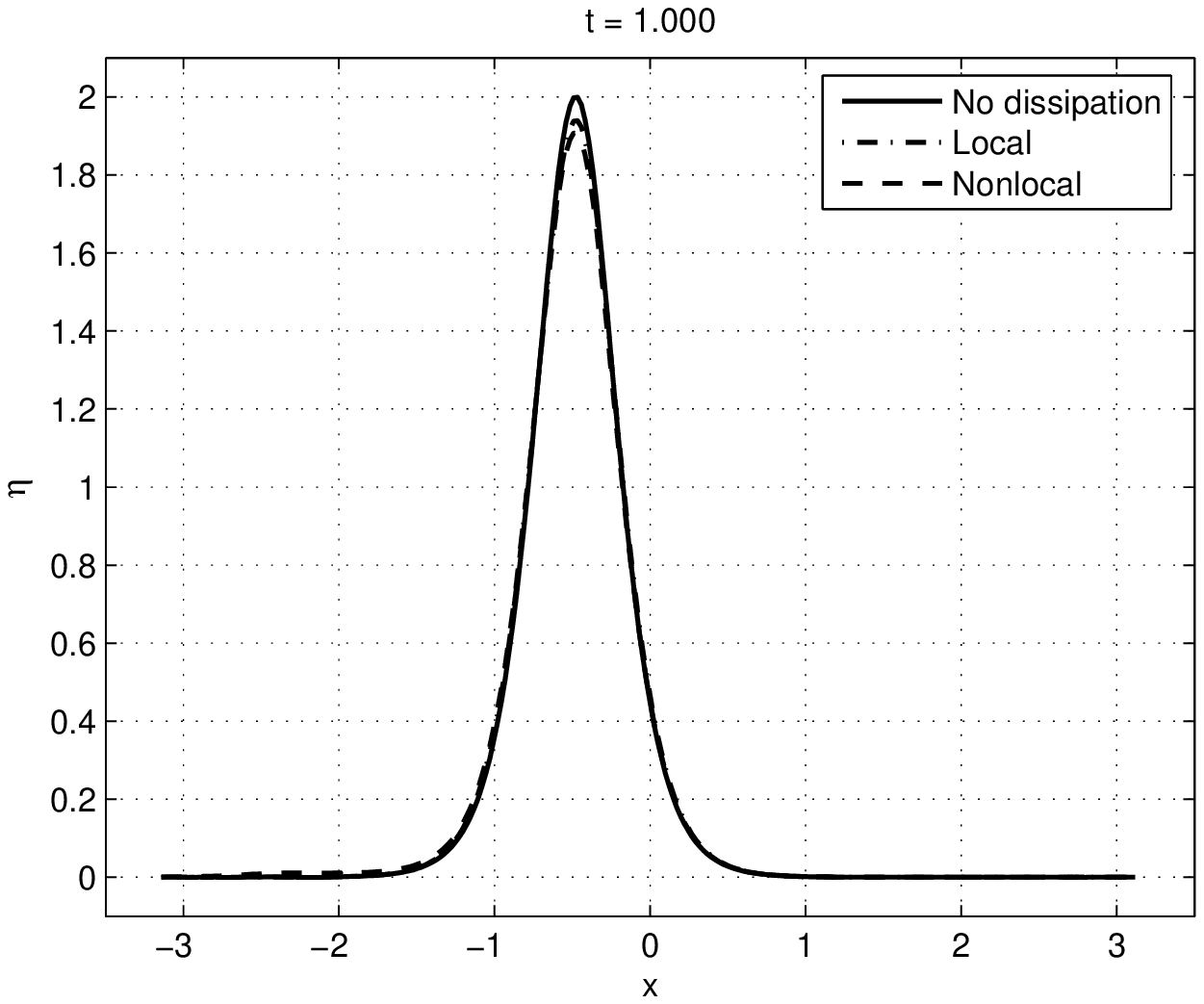}
	\caption[Snapshots of the free surface at $t=1.0$]{Comparison among two dissipative and nondissipative Boussinesq equations. Snapshots of the free surface at $t=1.0$}
	\label{fig:t1_0}
\end{figure}

\begin{figure}
	\centering
		\includegraphics[width=0.70\textwidth]{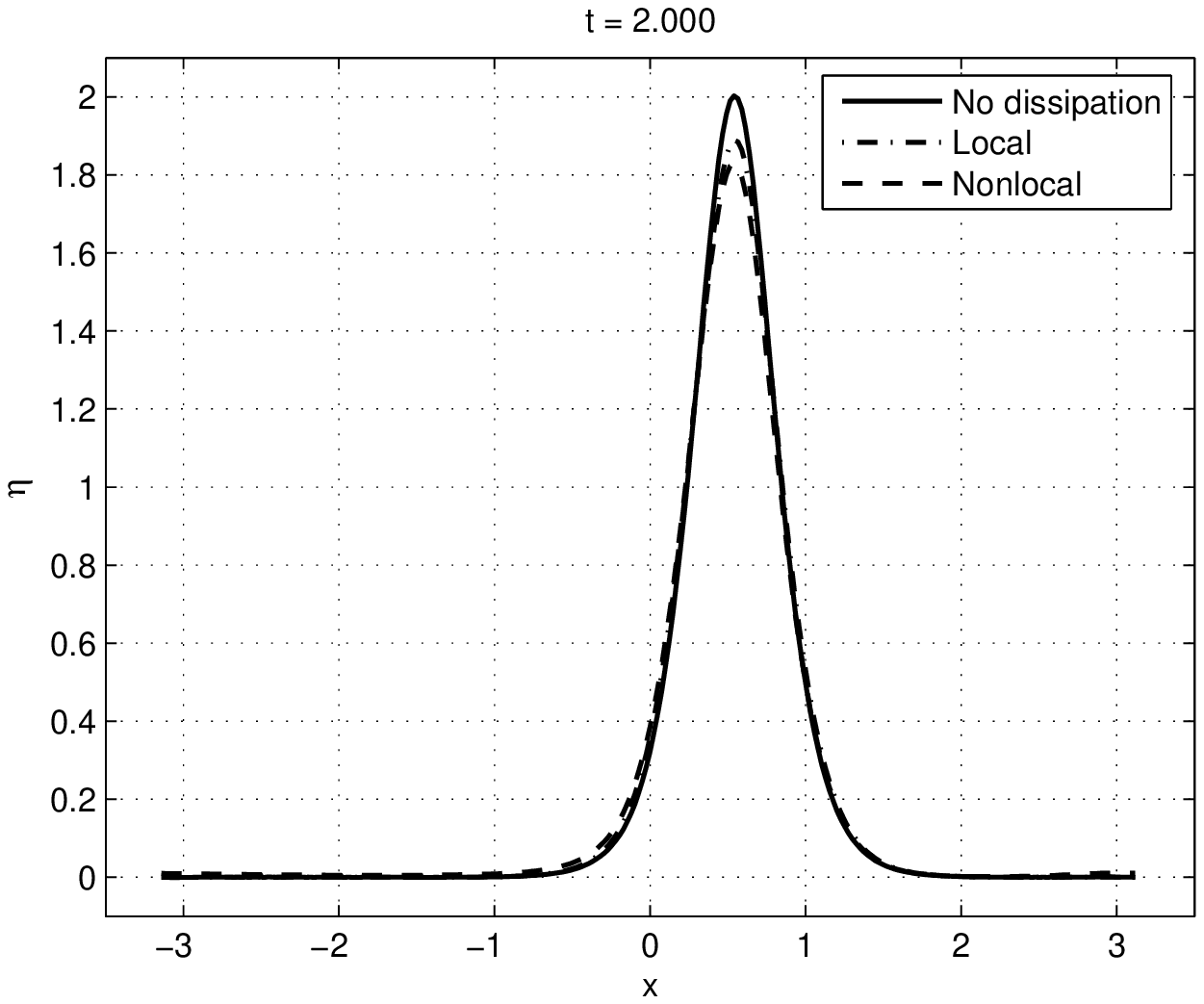}
	\caption[Snapshots of the free surface at $t=2.0$]{Comparison among two dissipative and nondissipative Boussinesq equations. Snapshots of the free surface at $t=2.0$}
	\label{fig:t2_0}
\end{figure}

\begin{figure}
	\centering
		\includegraphics[width=0.70\textwidth]{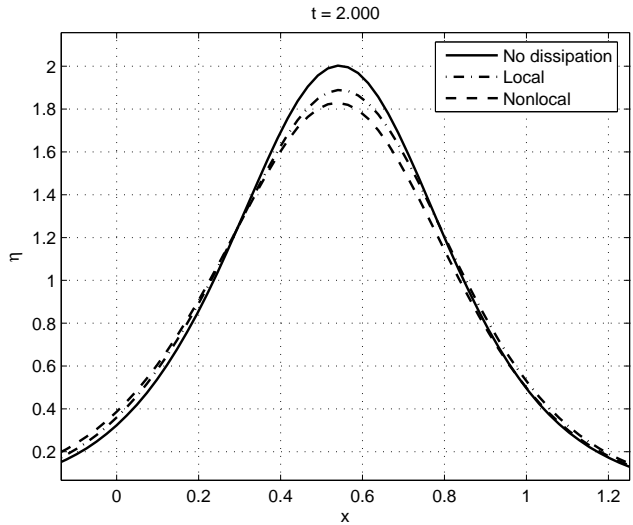}
	\caption[Zoom on the soliton crest at $t=2.0$]{Comparison among two dissipative and nondissipative Boussinesq equations. Zoom on the soliton crest at $t=2.0$}
	\label{fig:t2_0zoom}
\end{figure}

On Figures \ref{fig:t0_5}--\ref{fig:t2_0} we present three curves. They depict the free surface elevation according to three different formulations. The first corresponds to classical Boussinesq equations without dissipation. The second one to dissipative system with differential or local terms (for example, $\nu\Delta\u$ in momentum conservation equation) and the third curve corresponds to the new set of equations (\ref{bouss1}), (\ref{bouss2}). On \figurename~\ref{fig:t2_0zoom} we made a zoom on the soliton crest. 

It can be seen that Boussinesq equations with nonlocal term provide stronger attenuation of the amplitude. In the same time, as it was shown in the previous section, this nonlocal term slightly slows down the solitary wave.

\begin{figure}
	\centering
		\includegraphics[width=0.70\textwidth]{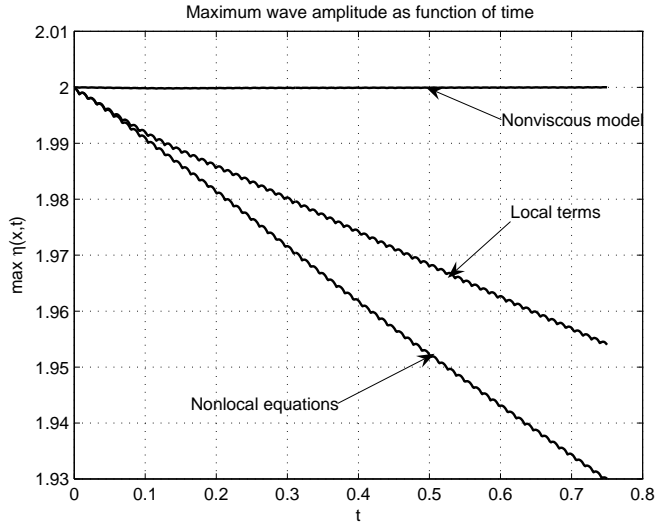}
	\caption{Amplitude of the solitary wave as function of dimensionless time.}
	\label{fig:amplitude}
\end{figure}

In order to show explicitly the rate of amplitude attenuation, we plot on \figurename~\ref{fig:amplitude} the graph of the following application $t\rightarrow \sup_{-\pi<x<\pi}|\eta(x,t)|$. One can see on this plot little oscillations which are of numerical nature. Our numerical experiments show that their amplitude decreases when $N\to\infty$. This result shows one more time that nonlocal model provides stronger damping properties. One can have the impression that the amplitude decays linearly but it is only an impression because of (\ref{eq:expdecay}). This behaviour for small $t$ can be explained by simple Taylor expansion which is valid when $\nu k^2 t\ll 1$:
\begin{equation*}
  \alpha(t)=\alpha_0 e^{-2\nu k^2 t} = \alpha_0(1 - 2\nu k^2 t) + 
  \O\bigl(\nu^2 k^4 t^2\bigr).
\end{equation*}

\section{Conclusion}



In the present article we used a novel visco-potential formulation proposed recently in \cite{Liu2004, DutykhDias2007, Dutykh2007a}. This formulation contains the nonlocal term in the kinematic bottom boundary condition. This term should be considered as a boundary layer correction at the bottom. In modelling viscous effects this term plays the main role, since its magnitude is $\O(\sqrt{\nu})$. Recently, important progress was made in numerical computation of the nonlocal term \cite{Torsvik2007}. Thus, we can hope to have its implementation in operational numerical codes. It is interesting to note that the boundary layer effect is not instantaneous but rather cumulative. The flow history is weighted by $(t-\tau)^{-\frac12}$ in favour of the current time $t$. As pointed out in \cite{Liu2004}, this nonlocal term is essential to have an accurate estimation of the bottom shear stress based on the calculated wave field above the bed. Then, this information can be used for calculating sediment-bedload transport fluxes and, in turn, morphological changes.

A long wave asymptotic limits (Boussinesq and KdV equations) were derived from this new formulation. The dispersion relation was described. Due to the presence of special functions, we cannot obtain a simple analytical dependence of the frequency $\omega$ on the wavenumber $k$ as in classical equations. Consequently the dispersion curve was obtained numerically by a Newton-type method. We made a comparison between the phase velocity of the complete visco-potential problem and the corresponding Boussinesq equations. The dispersion relation of new formulation is shown to be time dependent and this property is not classical. It comes from the memory effect of the boundary layer. We computed analytically the limiting dispersion curve $\omega_\infty(k)$ as $t\to+\infty$.

The effect of the nonlocal term on the solitary wave attenuation was investigated numerically with a Fourier-type spectral method. We showed that it provides much stronger damping (see \figurename~\ref{fig:amplitude}). An equation describing the amplitude evolution of linear progressive waves was derived (\ref{eq:at}). This result includes bottom boundary layer correction and generalizes classical formula (\ref{eq:classical0}) by Boussinesq \cite{Boussinesq1895} and Lamb \cite{Lamb1932}.

The present study opens new exciting possibilities for future research. Namely, we did not consider at all the questions of theoretical justification of visco-potential formulation and corresponding long wave models in the spirit of works by J. Bona, J.-C. Saut, D. Lannes, T. Colin \cite{BCL, BCS}. Other directions for future work consist in implementing new visco-potential framework in operational ocean and nearshore models.


\section*{Acknowledgment}
The author of this paper would like to express his gratitude to professors Fr\'{e}d\'{e}ric Dias and Jean-Claude Saut for very helpful discussions.


\bibliography{biblio}
\bibliographystyle{alpha}

\end{document}